\documentclass{pasa}%

\usepackage{graphicx}

\title[Spin direction asymmetry HST]{Galaxy spin direction distribution in HST and SDSS show similar large-scale asymmetry}

\author[Lior Shamir]{Lior Shamir
\affil{Kansas State University \\ Manhattan, KS 66506}%
}%

\jid{PASA}
\doi{10.1017/pas.\the\year.xxx}
\jyear{\the\year}

\usepackage{aas_macros}
\usepackage{hyperref} 
\hypersetup{colorlinks,citecolor=blue,linkcolor=blue,urlcolor=blue}

\hypersetup{draft}

\begin{document}

\begin{frontmatter}
\maketitle

\begin{abstract}
Several recent observations using large datasets of galaxies showed non-random distribution of the spin directions of spiral galaxies, even when the galaxies are too far from each other to have gravitational interaction. Here, a dataset of $\sim8.7\cdot10^3$ spiral galaxies imaged by Hubble Space Telescope is used to test and profile a possible asymmetry between galaxy spin directions. The asymmetry between galaxies with opposite spin directions is compared to the asymmetry of galaxies from the Sloan Digital Sky Survey. The two datasets contain different galaxies at different redshift ranges, and each dataset was annotated using a different annotation method. The results show that both datasets show a similar asymmetry in the COSMOS field, which is covered by both telescopes. Fitting the asymmetry of the galaxies to cosine dependence shows a dipole axis with probabilities of $\sim2.8\sigma$ and $\sim7.38\sigma$ in HST and SDSS, respectively. The most likely dipole axis identified in the HST galaxies is at $(\alpha=78^o,\delta=47^o)$, and is well within the $1\sigma$ error range compared to the location of the most likely dipole axis in the SDSS galaxies with $z>0.15$, identified at $(\alpha=71^o,\delta=61^o)$. 
\end{abstract}

\begin{keywords}
Galaxy: general -- galaxies: spiral
\end{keywords}
\end{frontmatter}

\section{INTRODUCTION}
\label{introduction}

Recently, several experiments using large datasets of galaxies imaged by several different instruments have shown evidence of non-random distribution of the spin directions of spiral galaxies  \citep{slosar2009galaxy,longo2011detection,shamir2012handedness,shamir2013color,hoehn2014characteristics,shamir2016asymmetry,shamir2017colour,shamir2017photometric,shamir2017large,lee2019galaxy,lee2019mysterious,shamir2019large,shamir2020asymmetry,shamir2020large,shamir2020patterns}.   The asymmetry is reflected by differences in the number of galaxies with opposite spin directions \citep{shamir2012handedness,shamir2019large,shamir2020patterns,lee2019mysterious}, and it changes with the directions of observation \citep{shamir2012handedness} and the redshift \citep{shamir2016asymmetry1,shamir2019large,shamir2020patterns}. Other experiments showed differences in the brightness of the galaxies \citep{shamir2016asymmetry,shamir2017large}.


Early experiments used galaxies annotated manually by a large number of volunteers showed no statistically significant difference between the number of galaxies with opposite spin directions \citep{land2008galaxy}. However, it was also found that volunteers annotating the same galaxies tended to classify elliptical galaxies with no apparent spin direction as spiral galaxies that spin clockwise, and therefore leading to a difference in the number of galaxies \citep{hayes2017nature}. Another experiment that used manual analysis of the data was based on five undergraduate students annotating $\sim1.5\cdot10^4$ galaxies. In that experiment the galaxies were also mirrored in attempt to correct for a possible human bias, and the results showed a difference of $\sim$7\% between the number of clockwise and counterclockwise galaxies \citep{longo2011detection}.

With the availability of very large astronomical databases, the ability to automate the annotation of the spin directions of spiral galaxies allowed to annotate far larger datasets. These large datasets of galaxies annotated by their spin direction can provide strong statistical signal, and profile a possible asymmetry between galaxies with opposite spin directions. It should be noted that the advantage of eliminating the human perception bias is compromised when using machine learning for the annotation, since machine learning algorithms are based on ``ground truth'' training data that is annotated manually, and the trained model can therefore still be biased by the data it was trained with.

By using model-driven automatic annotation algorithms \citep{shamir2011ganalyzer}, large datasets of galaxies showed asymmetry between the number of galaxies with opposite spin directions, and the asymmetry direction and magnitude change based on the direction of observation \citep{shamir2012handedness,shamir2019large,shamir2020large,shamir2020patterns} and the redshift \citep{shamir2016asymmetry1,shamir2019large,shamir2020patterns}. The asymmetry was identified in data collected by the Sloan Digital Sky Survey \citep{shamir2012handedness,shamir2016asymmetry}, and showed good agreement with the asymmetry identified in data collected by the Panoramic Survey Telescope and Rapid Response System \citep{shamir2017large,shamir2020asymmetry,shamir2020patterns}. 

Experiments with smaller datasets annotated manually also showed patterns of spin directions of galaxies \citep{slosar2009galaxy}, and alignment of spin directions was identified with quasars \citep{hutsemekers2014alignment}. More recently, consistency in spin directions was also observed with galaxies that are too distant from each other to have any kind of gravitational interactions \citep{lee2019mysterious}. These links are defined as ``mysterious'', leading to the assumption of a link between galaxy rotation and the motion of the large-scale structure \citep{lee2019mysterious}.


This paper shows an analysis of the asymmetry between galaxies with opposite spin directions observed when using spiral galaxies from different parts of the sky. The main dataset used in this study is taken from Hubble Space Telescope, and the asymmetry in that dataset is compared to the asymmetry in a galaxy dataset from SDSS used in previous experiments \citep{shamir2019large,shamir2020patterns}.

\section{DATA}
\label{data}

The dataset of spiral galaxies was taken from the Cosmic Assembly Near-infrared Deep Extragalactic Legacy Survey \citep{grogin2011candels,koekemoer2011candels}. The initial dataset contained 114,529 galaxies taken from the Great Observatories Origins Deep Survey North (GOODS-N), the Great Observatories Origins Deep Survey South (GOODS-S), the Ultra Deep Survey (UDS), the Extended Groth Strip (EGS), and the Cosmic Evolution Survey (COSMOS) fields. The galaxy images were separated from the F814W band FITS images using the {\it mSubimage} tool included in the {\it Montage} package \citep{berriman2004montage}, and were converted into 122$\times$122 TIF (Tagged Image File) images. 

The separation of the galaxies into galaxies with clockwise and counterclockwise spin directions was done manually. In previous experiments automatic analysis was used \citep{shamir2013color,shamir2017photometric,shamir2019large,shamir2020patterns}. However, while model-driven automatic analysis is unbiased and capable of analyzing very large databases, it is limited by its ability to classify all galaxies. Therefore, the spin direction of many galaxies cannot be determined, and these galaxies are excluded from the analysis. In sky surveys such as SDSS the number of galaxies is high, and therefore sacrificing some of the galaxies still leaves a sufficient number of accurately annotated galaxies, and does not affect the analysis as long as the algorithm is fully symmetric. However, the HST fields are far smaller than sky surveys such as SDSS, and sacrificing some of the galaxies can reduce the number of galaxies in the dataset. Another reason for using manual annotation is to use an accurate method that is different from the methods used in previous experiments.

\begin{figure}[h]
\centering
\includegraphics[scale=0.7]{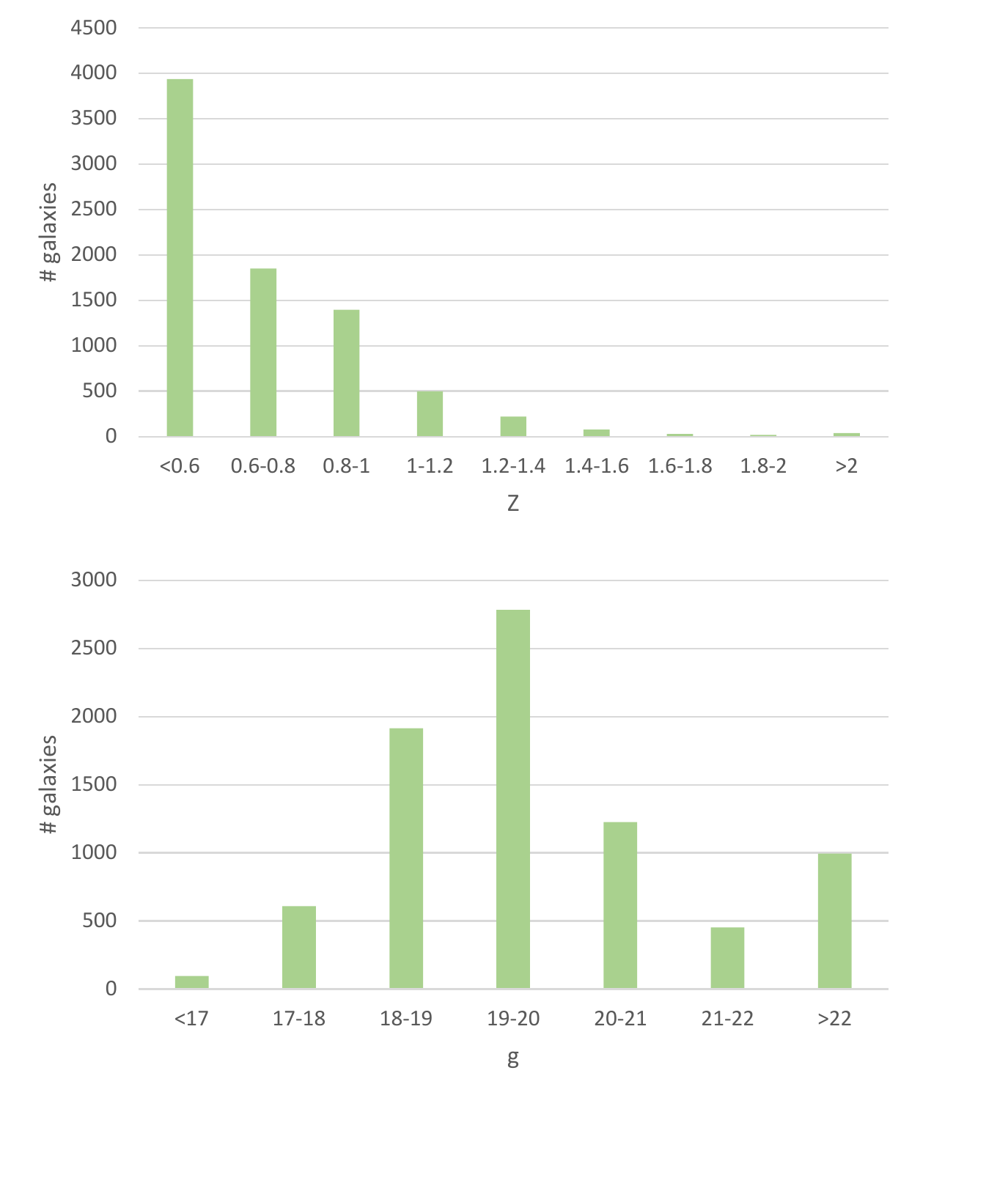}
\caption{The redshift and g magnitude distribution of the HST galaxies.}
\label{distribution}
\end{figure}

The annotation was done by first randomly mirroring half of the images, and then identifying all galaxies with clockwise spin direction and separating them from the rest of the galaxies. Then, all galaxy images were mirrored, and the clockwise galaxies were again separated from the rest of the galaxies. Each of these two datasets was then inspected to ensure that all galaxies are classified correctly. 
In the end of the process, 200 galaxies with clockwise spin direction, 200 galaxies with counterclockwise spin direction, and 200 galaxies that their spin direction could not be determined were inspected carefully. The examination showed that all 600 galaxies were annotated correctly. That provided a very clean dataset that is also symmetric in the annotations of the galaxies due to the random mirroring, and the identification of just clockwise galaxies. But unlike previous datasets, it is also complete in the sense that all galaxies that their spin direction could be determined are indeed annotated. The process was labor-intensive, and required $\sim$250 hours of work to complete. It provided a clean dataset of 8,690 galaxies with identifiable spin direction. The distribution of the galaxies in the different fields are summarized in Table~\ref{CANDELS}. The Subaru g magnitude and the photometric redshift distribution of these galaxies are shown in Figure~\ref{distribution}.

\begin{table*}
\caption{The number of galaxies in each of the five fields.}
\label{CANDELS}
\centering
\begin{tabular}{lccccc}
\hline
\hline
Field & Field                        & \# all       &  \# Clockwise   & \# Counterclockwise \\
        & center (degrees)       & galaxies   &  galaxies          & galaxies                  \\
\hline
GOODS-N & 189.23,62.24 & 5,931 & 396 & 373 \\
GOODS-S & 53.12,-27.81  & 5,024 & 276 & 264 \\
COSMOS &  150.12,2.2 & 84,424 & 3,116 & 2,965 \\
UDS        &  214.82,52.82           & 14,245 & 323 & 293 \\
EGS        &  34.41,-5.2              & 4,905 & 355 & 329 \\
\hline
\hline
\end{tabular}
\end{table*}

The distribution of spin directions in the HST galaxies was compared to datasets of SDSS and Pan-STARRS galaxies that were used in previous experiments \citep{shamir2017colour,shamir2017photometric,shamir2019large,shamir2020patterns}. These datasets were annotated automatically by the Ganalyzer \citep{shamir2011ganalyzer,ganalyzer_ascl} algorithm. Ganalyzer is a model-driven algorithm that is based on clear and defined rules. It is not based on machine learning or deep neural networks, and therefore cannot be biased by the training set or by complex non-intuitive rules typical to machine learning systems. In addition to the theoretical analysis of the algorithm, it also showed empirical evidence obtained by mirroring a large number of galaxy images. Full details about the galaxy annotation method can be found in \citep{shamir2017colour,shamir2017photometric,shamir2017large,shamir2020large}, and the dataset is described in \citep{shamir2020patterns}.

\section{RESULTS}
\label{results}


The distribution of galaxies in HST shows that the number of clockwise galaxies is higher, but the number of galaxies in the different fields is too low to allow statistical analysis. The only exception is the COSMOS field, where the number of galaxies is far higher than in any of the other HST fields used in this study. To compare the asymmetry in that field to galaxies imaged by SDSS and Pan-STARRS, the SDSS and Pan-STARRS galaxies in the 10$\times10$ degrees around the center of COSMOS were examined. The reason for using a larger field is because COSMOS is far deeper than SDSS and Pan-STARRS, and therefore SDSS and Pan-STARRS have a much smaller number of galaxies in a field of the same size. The difference between the size of the fields naturally makes the comparison indirect, as the fields being compared are different. But although the fields are not identical, such comparison can provide certain information regarding the agreement between the populations of galaxies in that part of the sky.

Datasets that were used in previous studies were examined, all of them were annotated automatically. These included a dataset of SDSS \citep{shamir2017photometric}, and a dataset of Pan-STARRS objects \citep{shamir2020patterns}. Because the dataset used in \citep{shamir2017photometric} contained photometric objects of extended sources, some of the photometric measurements were made from photometric objectss inside the same extended source. To avoid the presence of duplicate objects, all objects that had another object within 0.01$^o$ or less were removed. 
Detailed information about these datasets and the distribution of redshift and magnitude of the galaxies they contain are described in the relevant papers \citep{shamir2017photometric,shamir2017large,shamir2020patterns}. 

Table~\ref{cosmos_sdss} shows the number of galaxies by their spin directions in each of the three instruments. As the table shows, all datasets show a higher number of clockwise galaxies in that field. The statistical significance is not strong in the Pan-STARRS field, as expected due to the lower number of galaxies compared to COSMOS, but these fields do not conflict with the distribution of galaxy population in COSMOS. Assuming equal probability of having clockwise and counterclockwise galaxies, the probability of having that asymmetry in all of these fields is $2\times0.027\times0.017\times0.06\simeq5\cdot10^{-5}$.

\begin{table*}
\caption{Number of clockwise and counterclockwise galaxies in the COSMOS field and in the $10^o\times10^o$ field of SDSS and Pan-STARRS centered around COSMOS. The P value reflects the binomial probability of having asymmetry equal or greater than the observed asymmetry when assuming that a galaxy has 0.5 probability of having clockwise or counterclockwise spin direction. All of these datasets were annotated in an automatic process.}
\label{cosmos_sdss}
\centering
\begin{tabular}{lccc}
\hline
\hline
Survey &        \# Clockwise   & \# Counterclockwise & P \\
          &        galaxies          & galaxies                   & value \\
\hline
COSMOS &  3,116 & 2,965 & 0.027 \\
SDSS  \citep{shamir2017photometric}      &  350  & 295 & 0.017 \\
SDSS  \citep{shamir2020patterns}      &  461  & 440 & 0.24 \\
Pan-STARRS  \citep{shamir2020patterns}      &  222  & 190 & 0.06 \\
\hline
\hline
\end{tabular}
\end{table*}

Previous experiments showed evidence of non-random patterns of the asymmetry between the number of galaxies with opposite spin directions in different parts of the sky \citep{shamir2012handedness,shamir2019large,shamir2020patterns}. That was done by identifying the $(\alpha,\delta)$ at which the asymmetry of the galaxy spin directions had best fit to cosine dependence. The HST data used in this experiment include several different fields in different parts of the sky. That allows to fit the distribution of the spin directions of these galaxies to cosine dependence. Fitting the galaxy spin directions to cosine dependence can indicate whether the galaxy spin directions are aligned in a form of a possible dipole axis, and can also provide the statistical significance of such axis.

\begin{figure}[h]
\centering
\includegraphics[scale=0.15]{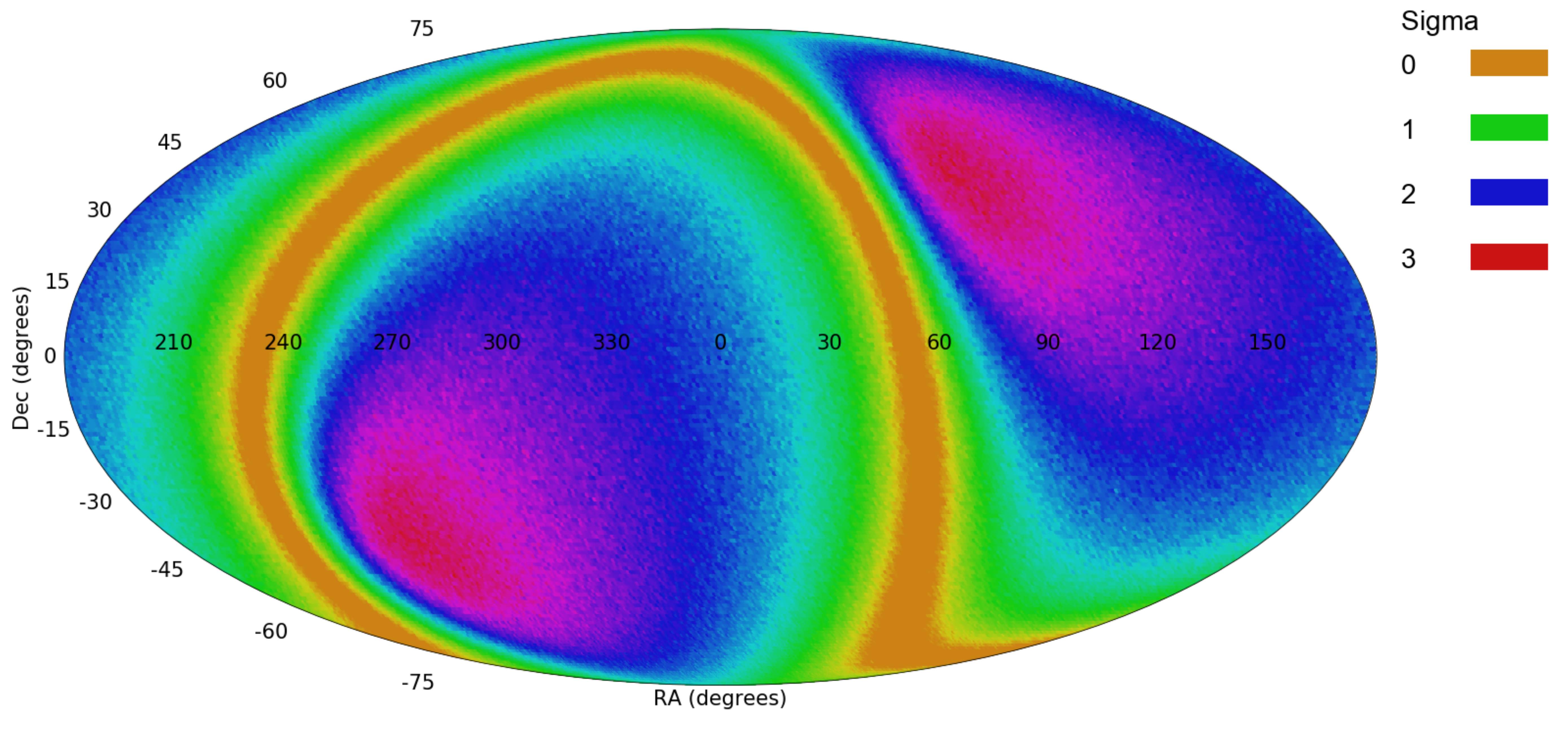}
\caption{Probability of cosine dependence of the spin directions of HST galaxies from every possible integer $(\alpha,\delta)$ combination.}
\label{hst_dipole}
\end{figure}

To test the probability that the spin direction asymmetry exhibits a dipole axis, the same method used in \citep{shamir2012handedness,shamir2019large,shamir2020patterns} was applied. Each galaxy was assigned with a value within the set $\{-1,1\}$. Galaxies with clockwise spin direction were assigned with 1, and galaxies with counterclockwise spin direction were assigned with -1. Then, $\chi^2$ statistics was used such that for each possible integer $(\alpha,\delta)$ combination, the angular distance $\phi$ between $(\alpha,\delta)$ and the celestial coordinates of each galaxy in the dataset was computed. The  $\cos(\phi)$ of the galaxies were then fitted into $d\cdot|\cos(\phi)|$, such that $d$ is the spin direction of the galaxy (a value within the set \{-1,1\}). The $\chi^2$ was computed 1000 times such that in each time the galaxies were assigned with random spin directions, and the mean and standard deviation were computed for each possible $(\alpha,\delta)$. The mean $\chi^2$ computed with the random spin directions was then compared to the $\chi^2$ computed when $d$ was assigned the real spin directions. The $\sigma$ difference between the $\chi^2$ of the real spin directions and the mean $\chi^2$ when using the random spin directions shows the likelihood of an axis at $(\alpha,\delta)$. When the likelihood of all $(\alpha,\delta)$ was computed, the $(\alpha,\delta)$ of the most likely dipole axis could be identified. Figure~\ref{hst_dipole} shows the probability of a dipole axis in all integer $(\alpha,\delta)$ combinations. The most likely axis was identified at $(\alpha=78^o,\delta=47^o)$, with probability of $\sim2.83\sigma$. The 1$\sigma$ error for that axis is $(58^0,184^o)$ for the right ascension, and $(6^o,73^o)$ for the declination. 

\begin{figure}[h]
\centering
\includegraphics[scale=0.27]{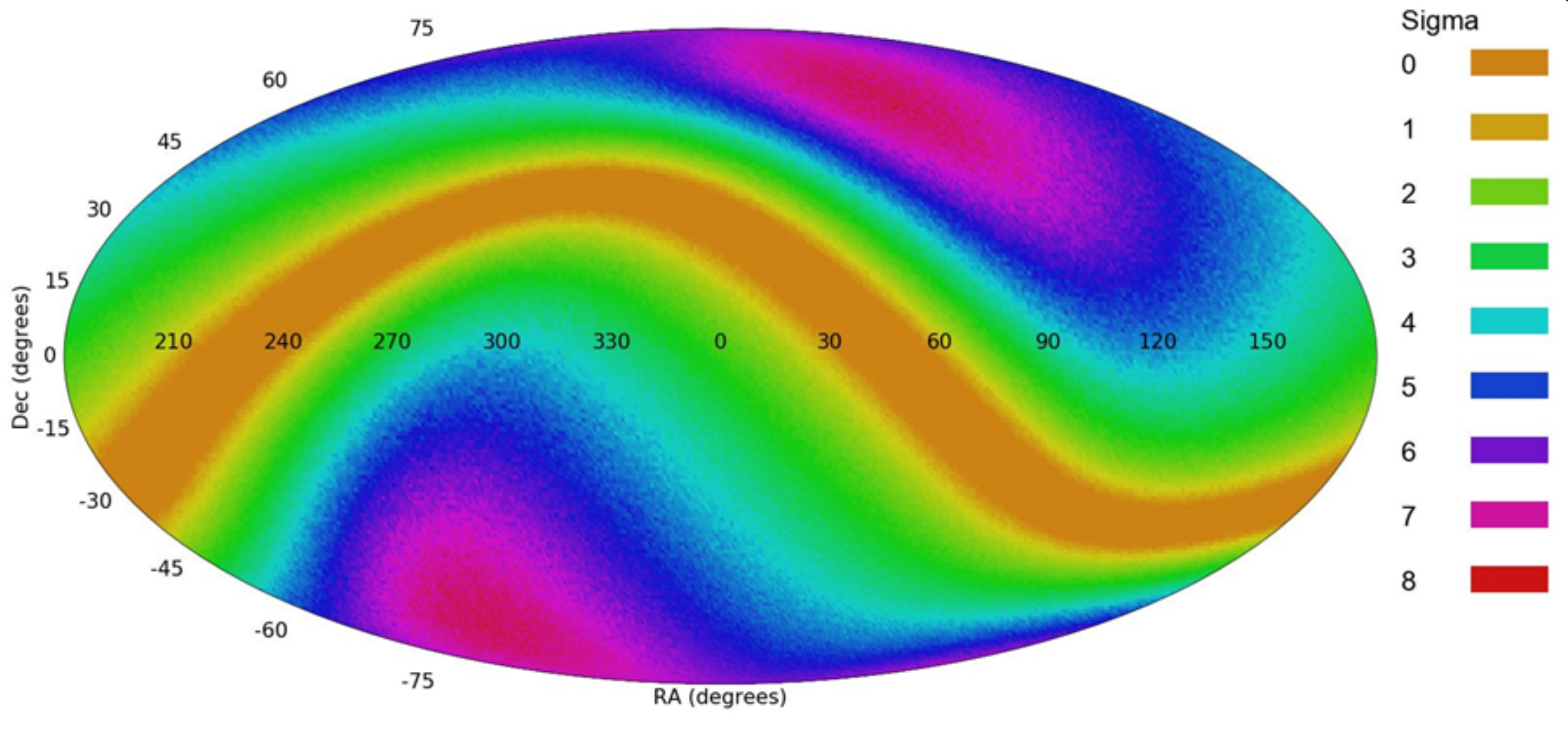}
\caption{Cosine dependence probability of the spin directions of SDSS galaxies from every possible integer $(\alpha,\delta)$ combination.}
\label{DR14_dipole_large}
\end{figure}

\begin{figure}[h]
\centering
\includegraphics[scale=0.3]{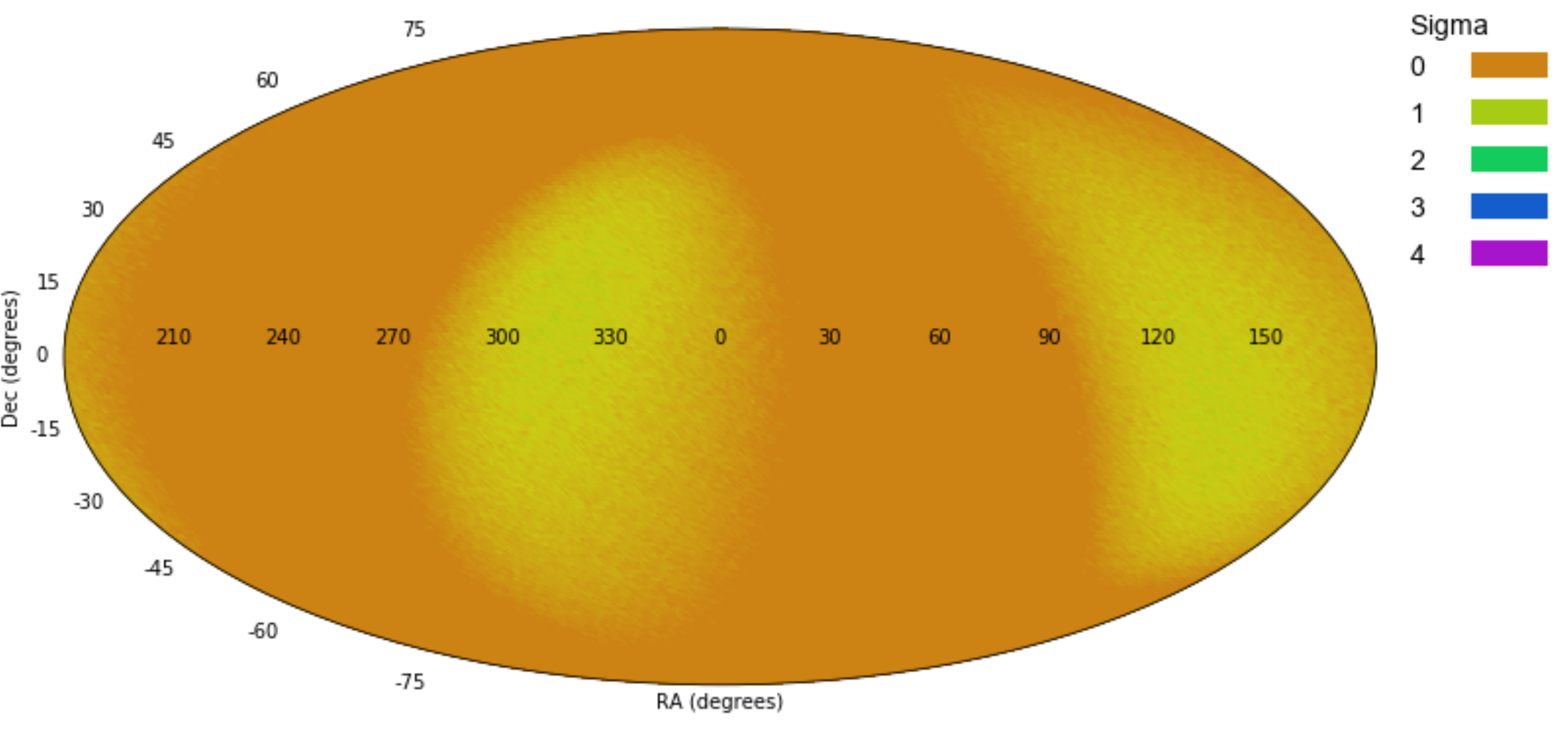}
\caption{Probability of cosine dependence of the spin directions of SDSS galaxies from every possible integer $(\alpha,\delta)$ combination when the galaxies are assigned with random spin directions.}
\label{DR14_dipole_large_random}
\end{figure}

The dipole axis identified in the HST galaxies was compared to the dipole axis identified in SDSS galaxies that were annotated automatically \citep{shamir2020patterns}. Figure~\ref{DR14_dipole_large} shows the probability of a dipole axis identified in each possible pair of integer $(\alpha,\delta)$ in the SDSS galaxies, when using the galaxies with $z>0.15$ used in \citep{shamir2020patterns}. That dataset included 15,863 galaxies annotated automatically by their spin direction. The most likely axis is identified at $(\alpha=71^o,\delta=61^o)$, with $\sigma\simeq7.38$. That most likely axis is close to the most likely dipole axis identified in the HST galaxies, and well within the 1$\sigma$ error. Figure~\ref{DR14_dipole_large_random} shows the most likely dipole axis when the galaxies are assigned with random spin directions. As expected, the dipole axis disappears when the galaxy spin directions are random.

\section{CONCLUSION}
\label{conclusion}

Results from two different datasets of galaxies imaged by two different instruments show similar asymmetry between galaxies with opposite spin directions. Each dataset contains different galaxies, and the galaxies in each dataset were annotated using a different method. Both datasets show a statistically significant dipole axis, and the location of the most likely axis is consistent in both datasets. Despite the difference in redshift, the two datasets show fairly similar location of the most likely dipole axis, and well within 1$\sigma$ error.
  
While the observations are clearly provocative, it is difficult to identify an error that could lead to such results. The experiments are based on two different instruments, and two different galaxy annotation methods. One of the instruments is space-based, reducing the possibility that the results are driven by an atmospheric effect. These results are consistent with previous similar experiments \citep{shamir2013color,shamir2016asymmetry,shamir2017colour,shamir2017photometric,shamir2017large,shamir2019large,shamir2020patterns}. The automatic annotation method is model-driven, does not rely on machine learning, and consistent when the galaxy images are mirrored \citep{shamir2017photometric}. Previous experiments also showed that the asymmetry changes in different parts of the sky, which is not expected if the annotation method is biased \citep{shamir2017large,shamir2019large,shamir2020patterns}.


It should be noted that while the vast majority of spiral galaxies are trailing, in some rare cases galaxies are counter-winding \citep{grouchy2008counter}. A small number of counter-winding galaxies can therefore lead to difference between the number of galaxies with opposite spin directions. However, if counter-winding galaxies are equally distributed between galaxies that spin clockwise and galaxies that spin counterclockwise, no statistically significant difference between the galaxies is expected. Therefore, to explain the observation with counter-winding galaxies, such galaxies need to have a certain preference based on the actual spin direction of the galaxy.

Analysis of the distribution of galaxies is limited by the fluctuations in large-scale galaxy population, known as ``cosmic variance'' \citep{moster2011cosmic}. However, here the measurement is a comparison between the number of galaxies with opposite spin directions identified inside the same exposures and same fields. It is therefore expected that fluctuations in galaxy population that affects the number of galaxies with a certain spin direction has the same impact on galaxies with the opposite spin direction. That reduces the possibility that the asymmetry is driven by changes in galaxy population, as any such change is expected to affect both clockwise and counterclockwise galaxies. This relative measurement is different from some other probes that are based on absolute measurements, such as the frequency of short GRBs or Ia supernovae. The use of a relative measurement can also handle effects such as Milky Way obstruction, as any obstruction that affects the ability to detect clockwise galaxies is expected to have a similar effect on the ability to detect counterclockwise galaxies in the same field.

It is naturally difficult to identify an immediate explanation for the observations. \cite{lee2019mysterious} identified consistency of spin directions of galaxies even if the galaxies are too far to interact gravitationally, and defined the observation as ``mysterious'' \citep{lee2019mysterious}. Explanations of the asymmetry can be related to parity-breaking gravitational waves, which can affect galaxy shape during inflation \citep{biagetti2020primordial}, and can provide an explanation to the asymmetry without violating the basic cosmological assumptions. Cosmological-scale anisotropy has been observed in the past with cosmic microwave background \citep{cline2003does,gordon2004low,zhe2015quadrupole}. These observations also challenge the basic cosmological assumptions and led to theories that differ from the standard cosmological models \citep{feng2003double,piao2004suppressing,rodrigues2008anisotropic,piao2005possible,jimenez2007cosmology,bohmer2008cmb}. These observations also led to the model of ellipsoidal universe \citep{campanelli2006ellipsoidal,campanelli2007cosmic,gruppuso2007complete}, as well as a rotating universe \citep{godel1949example,ozsvath1962finite,ozsvath2001approaches,sivaram2012primordial,chechin2016rotation}.

Cosmological isotropy and homogeneity are basic assumptions used in most standard cosmological theories, although spatial homogeneity is an assumption that cannot be verified directly \citep{ellis1979homogeneity}. Some evidence of cosmological isotropy violation have been observed through other messengers such as radio sources \citep{bengaly2018probing}, luminosity-temperature ratio \citep{migkas2020probing}, short gamma ray bursts \citep{meszaros2019oppositeness}, Ia supernova \citep{javanmardi2015probing}, distribution of galaxy morphology \citep{javanmardi2017anisotropy}, and cosmic microwave background \citep{aghanim2014planck,hu1997cmb,cooray2003cosmic,ben2012parity,eriksen2004asymmetries}. Future instruments such as the Earth-based Rubin observatory and the space-based Euclid can be used to validate whether the asymmetry is observed also in other instruments, and provide better profiling of the asymmetry.

Given the multiple reports on anomaly in the distribution of galaxies with opposite spin patterns \citep{longo2011detection,shamir2012handedness,shamir2019large,lee2019mysterious,shamir2020patterns}, it is important to continue the examination of such observations, verifying and profiling the distribution, and identifying whether the reported observations can have non-astronomical explanations.

\section*{ACKNOWLEDGMENTS}

I would like to thank the anonymous reviewer for the insightful comments. This study was supported in part by NSF grants AST-1903823 and IIS-1546079.

The research is based on observations made with the NASA/ESA Hubble Space Telescope, and obtained from the Hubble Legacy Archive, which is a collaboration between the Space Telescope Science Institute (STScI/NASA), the Space Telescope European Coordinating Facility (ST-ECF/ESA) and the Canadian Astronomy Data Centre (CADC/NRC/CSA).

SDSS-IV is managed by the Astrophysical Research Consortium for the Participating Institutions of the SDSS Collaboration including the Brazilian Participation Group, the Carnegie Institution for Science, Carnegie Mellon University, the Chilean Participation Group, the French Participation Group, Harvard-Smithsonian Center for Astrophysics, Instituto de Astrofisica de Canarias, The Johns Hopkins University, Kavli Institute for the Physics and Mathematics of the Universe (IPMU) / 
University of Tokyo, the Korean Participation Group, Lawrence Berkeley National Laboratory, Leibniz Institut fur Astrophysik Potsdam (AIP), Max-Planck-Institut fur Astronomie (MPIA Heidelberg), Max-Planck-Institut fur Astrophysik (MPA Garching), Max-Planck-Institut fur Extraterrestrische Physik (MPE), National Astronomical Observatories of China, New Mexico State University, New York University, University of Notre Dame, Observatario Nacional / MCTI, The Ohio State University, Pennsylvania State University, Shanghai Astronomical Observatory, United Kingdom Participation Group, Universidad Nacional Autonoma de Mexico, University of Arizona, University of Colorado Boulder, University of Oxford, University of Portsmouth, University of Utah, University of Virginia, University of Washington, University of Wisconsin, Vanderbilt University, and Yale University.


\begin{thebibliography}{}
\makeatletter
\relax
\def\mn@urlcharsother{\let\do\@makeother \do\$\do\&\do\#\do\^\do\_\do\%\do\~}
\definecolor{darkblue}{rgb}{0,0,0.597656}
\def\mndoi{\begingroup\mn@urlcharsother \@ifnextchar [ {\mndoi@} {\mndoi@[]}}
\def\mndoi@[#1]#2{\def\@tempa{#1}\ifx\@tempa\@empty \href
  {http://dx.doi.org/#2} {\textcolor{darkblue}{doi:#2}}\else \href
  {http://dx.doi.org/#2} {\textcolor{darkblue}{#1}}\fi \endgroup}
\def\mn@eprint#1#2{\mn@eprint@#1:#2::\@nil}
\def\mn@eprint@arXiv#1{\href {http://arxiv.org/abs/#1} {{\tt arXiv:#1}}}
\def\mn@eprint@dblp#1{\href {http://dblp.uni-trier.de/rec/bibtex/#1.xml}
  {dblp:#1}}
\def\mn@eprint@#1:#2:#3:#4\@nil{\def\@tempa {#1}\def\@tempb {#2}\def\@tempc
  {#3}\ifx \@tempc \@empty \let \@tempc \@tempb \let \@tempb \@tempa \fi \ifx
  \@tempb \@empty \def\@tempb {arXiv}\fi \@ifundefined
  {mn@eprint@\@tempb}{\@tempb:\@tempc}{\expandafter \expandafter \csname
  mn@eprint@\@tempb\endcsname \expandafter{\@tempc}}}

\bibitem[\protect\citeauthoryear{Aghanim et~al.,}{Aghanim
  et~al.}{2014}]{aghanim2014planck}
Aghanim N.,  et~al., 2014, A\&A, 571, A27

\bibitem[\protect\citeauthoryear{Ben-David, Kovetz  \& Itzhaki}{Ben-David
  et~al.}{2012}]{ben2012parity}
Ben-David A.,  Kovetz E.~D.,   Itzhaki N.,  2012, ApJ, 748, 39

\bibitem[\protect\citeauthoryear{Bengaly, Maartens  \& Santos}{Bengaly
  et~al.}{2018}]{bengaly2018probing}
Bengaly C.~A.,  Maartens R.,   Santos M.~G.,  2018, Journal of Cosmology and
  Astroparticle Physics, 2018, 031

\bibitem[\protect\citeauthoryear{Berriman et~al.,}{Berriman
  et~al.}{2004}]{berriman2004montage}
Berriman G.~B.,  et~al., 2004, in ADASS XIII. p.~593

\bibitem[\protect\citeauthoryear{Biagetti \& Orlando}{Biagetti \&
  Orlando}{2020}]{biagetti2020primordial}
Biagetti M.,  Orlando G.,  2020, Journal of Cosmology and Astroparticle
  Physics, 2020

\bibitem[\protect\citeauthoryear{Bohmer \& Mota}{Bohmer \&
  Mota}{2008}]{bohmer2008cmb}
Bohmer C.~G.,  Mota D.~F.,  2008, PLB, 663, 168

\bibitem[\protect\citeauthoryear{Campanelli, Cea  \& Tedesco}{Campanelli
  et~al.}{2006}]{campanelli2006ellipsoidal}
Campanelli L.,  Cea P.,   Tedesco L.,  2006, PRL, 97, 131302

\bibitem[\protect\citeauthoryear{Campanelli, Cea  \& Tedesco}{Campanelli
  et~al.}{2007}]{campanelli2007cosmic}
Campanelli L.,  Cea P.,   Tedesco L.,  2007, PRD, 76, 063007

\bibitem[\protect\citeauthoryear{Chechin}{Chechin}{2016}]{chechin2016rotation}
Chechin L.,  2016, Astronomy Reports, 60, 535

\bibitem[\protect\citeauthoryear{Cline, Crotty  \& Lesgourgues}{Cline
  et~al.}{2003}]{cline2003does}
Cline J.~M.,  Crotty P.,   Lesgourgues J.,  2003, Journal of Cosmology and
  Astroparticle Physics, 2003, 010

\bibitem[\protect\citeauthoryear{Cooray, Melchiorri  \& Silk}{Cooray
  et~al.}{2003}]{cooray2003cosmic}
Cooray A.,  Melchiorri A.,   Silk J.,  2003, PLB, 554, 1

\bibitem[\protect\citeauthoryear{Ellis}{Ellis}{1979}]{ellis1979homogeneity}
Ellis G.,  1979, General Relativity and Gravitation, 11, 281

\bibitem[\protect\citeauthoryear{Eriksen, Hansen, Banday, Gorski  \&
  Lilje}{Eriksen et~al.}{2004}]{eriksen2004asymmetries}
Eriksen H.~K.,  Hansen F.~K.,  Banday A.~J.,  Gorski K.~M.,   Lilje P.~B.,
  2004, ApJ, 605, 14

\bibitem[\protect\citeauthoryear{Feng \& Zhang}{Feng \&
  Zhang}{2003}]{feng2003double}
Feng B.,  Zhang X.,  2003, PLB, 570, 145

\bibitem[\protect\citeauthoryear{G{\"o}del}{G{\"o}del}{1949}]{godel1949example}
G{\"o}del K.,  1949, Reviews of Modern Physics, 21, 447

\bibitem[\protect\citeauthoryear{Gordon \& Hu}{Gordon \&
  Hu}{2004}]{gordon2004low}
Gordon C.,  Hu W.,  2004, PRD, 70, 083003

\bibitem[\protect\citeauthoryear{Grogin et~al.,}{Grogin
  et~al.}{2011}]{grogin2011candels}
Grogin N.~A.,  et~al., 2011, ApJS, 197, 35

\bibitem[\protect\citeauthoryear{Grouchy et~al.,}{Grouchy
  et~al.}{2008}]{grouchy2008counter}
Grouchy R.~D., Buta, R., Salo, H., Laurikainen, E., Speltincx, T., 2011, AJ, 136, 3

\bibitem[\protect\citeauthoryear{Gruppuso}{Gruppuso}{2007}]{gruppuso2007complete}
Gruppuso A.,  2007, PRD, 76, 083010

\bibitem[\protect\citeauthoryear{Hayes, Davis  \& Silva}{Hayes
  et~al.}{2017}]{hayes2017nature}
Hayes W.~B.,  Davis D.,   Silva P.,  2017, MNRAS, 466, 3928

\bibitem[\protect\citeauthoryear{Hoehn \& Shamir}{Hoehn \&
  Shamir}{2014}]{hoehn2014characteristics}
Hoehn C.,  Shamir L.,  2014, AN, 335, 189

\bibitem[\protect\citeauthoryear{Hu \& White}{Hu \& White}{1997}]{hu1997cmb}
Hu W.,  White M.,  1997, arXiv preprint astro-ph/9706147

\bibitem[\protect\citeauthoryear{Hutsem{\'e}kers, Braibant, Pelgrims  \&
  Sluse}{Hutsem{\'e}kers et~al.}{2014}]{hutsemekers2014alignment}
Hutsem{\'e}kers D.,  Braibant L.,  Pelgrims V.,   Sluse D.,  2014, A\&A, 572,
  A18

\bibitem[\protect\citeauthoryear{Javanmardi \& Kroupa}{Javanmardi \&
  Kroupa}{2017}]{javanmardi2017anisotropy}
Javanmardi B.,  Kroupa P.,  2017, A\&A, 597, A120

\bibitem[\protect\citeauthoryear{Javanmardi, Porciani, Kroupa  \&
  Pflam-Altenburg}{Javanmardi et~al.}{2015}]{javanmardi2015probing}
Javanmardi B.,  Porciani C.,  Kroupa P.,   Pflam-Altenburg J.,  2015, ApJ, 810,
  47

\bibitem[\protect\citeauthoryear{Jim{\'e}nez \& Maroto}{Jim{\'e}nez \&
  Maroto}{2007}]{jimenez2007cosmology}
Jim{\'e}nez J.~B.,  Maroto A.~L.,  2007, PRD, 76, 023003

\bibitem[\protect\citeauthoryear{Koekemoer et~al.,}{Koekemoer
  et~al.}{2011}]{koekemoer2011candels}
Koekemoer A.~M.,  et~al., 2011, ApJS, 197, 36

\bibitem[\protect\citeauthoryear{Land et~al.,}{Land
  et~al.}{2008}]{land2008galaxy}
Land K.,  et~al., 2008, MNRAS, 388, 1686

\bibitem[\protect\citeauthoryear{Lee, Pak, Lee  \& Song}{Lee
  et~al.}{2019a}]{lee2019galaxy}
Lee J.~H.,  Pak M.,  Lee H.-R.,   Song H.,  2019a, ApJ, 872, 78

\bibitem[\protect\citeauthoryear{Lee, Pak, Song, Lee, Kim  \& Jeong}{Lee
  et~al.}{2019b}]{lee2019mysterious}
Lee J.~H.,  Pak M.,  Song H.,  Lee H.-R.,  Kim S.,   Jeong H.,  2019b, ApJ,
  884, 104

\bibitem[\protect\citeauthoryear{Longo}{Longo}{2011}]{longo2011detection}
Longo M.~J.,  2011, PLB, 699, 224

\bibitem[\protect\citeauthoryear{M{\'e}sz{\'a}ros}{M{\'e}sz{\'a}ros}{2019}]{meszaros2019oppositeness}
M{\'e}sz{\'a}ros A.,  2019, AN, 340, 564

\bibitem[\protect\citeauthoryear{Migkas, Schellenberger, Reiprich, Pacaud,
  Ramos-Ceja  \& Lovisari}{Migkas et~al.}{2020}]{migkas2020probing}
Migkas K.,  Schellenberger G.,  Reiprich T.,  Pacaud F.,  Ramos-Ceja M.,
  Lovisari L.,  2020, A\&A, 636, A15

\bibitem[\protect\citeauthoryear{Moster et~al.,}{Moster
  et~al.}{2011}]{moster2011cosmic}
  Moster, Benjamin P and Somerville, Rachel S and Newman, Jeffrey A and Rix, Hans-Walter, 2011, ApJ, 731, 113 

\bibitem[\protect\citeauthoryear{Ozsv{\'a}th \& Sch{\"u}cking}{Ozsv{\'a}th \&
  Sch{\"u}cking}{1962}]{ozsvath1962finite}
Ozsv{\'a}th I.,  Sch{\"u}cking E.,  1962, Nature, 193, 1168

\bibitem[\protect\citeauthoryear{Ozsvath \& Sch{\"u}cking}{Ozsvath \&
  Sch{\"u}cking}{2001}]{ozsvath2001approaches}
Ozsvath I.,  Sch{\"u}cking E.,  2001, Classical and Quantum Gravity, 18, 2243

\bibitem[\protect\citeauthoryear{Piao}{Piao}{2005}]{piao2005possible}
Piao Y.-S.,  2005, PRD, 71, 087301

\bibitem[\protect\citeauthoryear{Piao, Feng  \& Zhang}{Piao
  et~al.}{2004}]{piao2004suppressing}
Piao Y.-S.,  Feng B.,   Zhang X.,  2004, PRD, 69, 103520

\bibitem[\protect\citeauthoryear{Rodrigues}{Rodrigues}{2008}]{rodrigues2008anisotropic}
Rodrigues D.~C.,  2008, PRD, 77, 023534

\bibitem[\protect\citeauthoryear{Shamir}{Shamir}{2011a}]{ganalyzer_ascl}
Shamir L.,  2011a, The Astrophysics Source Code Library, p. ascl:1105.011

\bibitem[\protect\citeauthoryear{Shamir}{Shamir}{2011b}]{shamir2011ganalyzer}
Shamir L.,  2011b, ApJ, 736, 141

\bibitem[\protect\citeauthoryear{Shamir}{Shamir}{2012}]{shamir2012handedness}
Shamir L.,  2012, PLB, 715, 25

\bibitem[\protect\citeauthoryear{Shamir}{Shamir}{2013}]{shamir2013color}
Shamir L.,  2013, Galaxies, 1, 210

\bibitem[\protect\citeauthoryear{Shamir}{Shamir}{2016a}]{shamir2016asymmetry1}
Shamir L.,  2016a, arXiv:1601.04424v1

\bibitem[\protect\citeauthoryear{Shamir}{Shamir}{2016b}]{shamir2016asymmetry}
Shamir L.,  2016b, ApJ, 823, 32

\bibitem[\protect\citeauthoryear{Shamir}{Shamir}{2017a}]{shamir2017large}
Shamir L.,  2017a, PASA, 34, e44

\bibitem[\protect\citeauthoryear{Shamir}{Shamir}{2017b}]{shamir2017photometric}
Shamir L.,  2017b, PASA, 34, e011

\bibitem[\protect\citeauthoryear{Shamir}{Shamir}{2017c}]{shamir2017colour}
Shamir L.,  2017c, ApSS, 362, 33

\bibitem[\protect\citeauthoryear{Shamir}{Shamir}{2019}]{shamir2019large}
Shamir L.,  2019, arXiv, p. 1912.05429

\bibitem[\protect\citeauthoryear{Shamir}{Shamir}{2020a}]{shamir2020asymmetry}
Shamir L.,  2020a, Open Astronomy, 29, 15

\bibitem[\protect\citeauthoryear{Shamir}{Shamir}{2020b}]{shamir2020large}
Shamir L.,  2020b, AAS, 236, 336.02

\bibitem[\protect\citeauthoryear{Shamir}{Shamir}{2020c}]{shamir2020patterns}
Shamir L.,  2020c, ApSS, 365, 136

\bibitem[\protect\citeauthoryear{Sivaram \& Arun}{Sivaram \&
  Arun}{2012}]{sivaram2012primordial}
Sivaram C.,  Arun K.,  2012, Open Astronomy, 5, 7

\bibitem[\protect\citeauthoryear{Slosar et~al.,}{Slosar
  et~al.}{2009}]{slosar2009galaxy}
Slosar A.,  et~al., 2009, MNRAS, 392, 1225

\bibitem[\protect\citeauthoryear{Zhe, Xin  \& Sai}{Zhe
  et~al.}{2015}]{zhe2015quadrupole}
Zhe C.,  Xin L.,   Sai W.,  2015, Chinese Physics C, 39, 055101

\makeatother
\end{thebibliography}

\end{document}